\documentclass[aps, preprint]{revtex4-1}
\draft % marks overfull lines with a black rule on the right

\usepackage{dcolumn}
\usepackage{amsbsy}
\usepackage{amsmath}
\usepackage{subfigure}
\usepackage{amssymb}
\usepackage{braket}
\usepackage{color,xcolor}
\usepackage{fullpage}
\usepackage{graphicx}
\usepackage{caption}
\usepackage{bm}
\usepackage{cleveref}
\usepackage{comment}
\usepackage{epstopdf}
\usepackage{mathrsfs}
\usepackage{upgreek}

\usepackage{blindtext}
\usepackage{mathtools}
\usepackage[justification=justified,singlelinecheck=false]{caption}

\newcommand{\bd}{\mathbf{d}}

\newcommand{\bmm}{\mathbf{m}}

\newcommand{\br}{\mathbf{r}}

\newcommand{\bF}{\mathbf{F}}

\newcommand{\bK}{\mathbf{K}}

\newcommand{\btheta}{\boldsymbol{\theta}}
\newcommand{\bTheta}{\boldsymbol{\Theta}}

\begin{document}

%\preprint{}

%\preprint{Submitted to Physical Review X}

\title{The role of nonlinearities in topological protection: using magnetically coupled fidget spinners}
\author{Javier Vila$^{1,2}$, Glaucio H. Paulino$^1$, Massimo Ruzzene$^{2,3}$}
%{\small $^b$ School of Mechanical Engineering, Georgia Institute of Technology, Atlanta GA 30332}\\
%{\small$^*$Corresponding author. E-mail: raj.pal@aerospace.gatech.edu}}
\email{Corresponding author: ruzzene@gatech.edu}
\affiliation{$^1$ School of Civil Engineering, Georgia Institute of Technology, Atlanta GA 30332}
\affiliation{$^2$ School of Aerospace Engineering, Georgia Institute of Technology, Atlanta GA 30332}
\affiliation{$^3$ School of Mechanical Engineering, Georgia Institute of Technology, Atlanta GA 30332}

\date{\today}

\begin{abstract}
	We investigate and experimentally observe the existence of topologically protected interface modes in a one-dimensional mechanical lattice, and we report on the effect of nonlinearities on topological protection. The lattice consists of a one-dimensional array of spinners with nearest neighbor coupling resulting from magnetic interactions. The distance between the spinners is spatially modulated to obtain a diatomic configuration, and to produce a non-trivial interface by breaking spatial inversion symmetry. For small amplitudes of motion, the interactions are approximately linear, and the system supports topologically protected interface modes at frequencies inside the bulk bandgaps of the lattice. Nonlinearities induced by increasing amplitude of motion cause the interface modes to shift outside the bandgaps and merge with the bulk bands. The resulting edge-to-bulk transition causes the extinction of the topologically protected interface mode and extends it to the entire length of the chain. Such transition is predicted by analytical calculations and verified by experimental observations. The paper thus investigates the existence of topologically protected interface modes obtained through broken spatial inversion symmetry, and documents their lack of robustness in the presence of nonlinearities.
\end{abstract}

% \pacs{}% insert suggested PACS numbers in braces on next line

\maketitle %\maketitle must follow title, authors, abstract and \pacs


\section{Introduction}\label{Sec. Intro}

Notable efforts have been devoted to the investigation of topological protection in condensed matter~\cite{Kane2005a,Kane2005b}, and in classical areas of wave physics such as acoustics \cite{Brendel2018}, photonics \cite{Noh2018,Ni2018}, as well as solid~\cite{Pal2017a,Vila2017PRB} and fluid mechanics~\cite{Banerjee2017}. The phenomenon of topological protection consists in the existence of wave modes that do not propagate into the bulk of the considered media, but are instead confined to a lower dimensional region within it, either a boundary or an interface. Driven by its topological nature, this effect is robust to the existence of imperfections and defects, making it attractive for applications where lossless wave propagation, immunity to backscattering and mode localization are important objectives. Topological protection can be achieved through time-reversal symmetry breaking, which generally requires the employment of active elements that effectively bias the interactions within the media. Examples include circulators in optomechanics \cite{Ruesink2018}, gyroscopic mechanical metamaterial~\cite{Noah2018}, and the use of active fluids characterized by a background flow~\cite{Souslov2017} among others. These systems mimic the quantum Hall effect whereby a net ``magnetic" flow breaks time-reversal symmetry. Two superimposing effects lead to the emergence of topologically protected (TP) modes. First, bandgaps are opened at the otherwise high symmetry degeneracy points in reciprocal space (e.g. Dirac cones). Second, the integral of the Berry curvature of each band along the reciprocal space does not equal zero, and the separated dispersion surfaces are linked only by one lower dimensional band, which corresponds to the TP mode. The eigenvector associated with this band is localized to a lower dimensional region in space and propagation is unidirectional \cite{Hatsugai1997}. Similar effects are achieved through solely passive elements that break spatial inversion symmetry (SIS)~\cite{Pal2016JAP,Wu2018}. Spatial inversion symmetry breaking also opens bandgaps at the high symmetry points and couples the spins (or polarities) of otherwise degenerate modes. In this case, the integral of the Berry curvature is nonzero in the vicinity of the opened Dirac cone, although it is zero over the entire reciprocal space. Two lower dimensional bands are produced and are associated with TP modes localized at the interface of two lattices with inverted bands, i.e. bands that are characterized by opposite values of the relevant topological invariants, and propagation of these associated modes occurs in opposite directions~\cite{Vila2017PRB}.

In systems that involve active elements, topological protection may be tailored or removed by control of such elements. In passive systems, the control of TP modes must instead rely on the inherent dynamic behavior of the lattice. Thus, nonlinearities appear as natural choices to pursue the objective of controlling and tailoring TP modes. Indeed, the vast majority of studies in the field of topological protection is limited to linear systems. While some theoretical investigations involving topological transitions have been recently presented~\cite{Hadad2016PRB,Pal2018PRE}, the physical demonstration of how nonlinearities affect TP modes remains mostly unexplored. Nonlinearities, for example, enable uneven distributions of the wave energy, which in turn may lead to nonreciprocal wave propagation \cite{VILA2017JSV,Luo2018,Moore2018,KUMAR2017,Raney2016}. Another interesting nonlinear effect is the change in the effective parameters governing wave motion, such as the equivalent stiffness of elastic systems, which produces shifts of dispersion branches and bandgaps~\cite{Leamy:2010mw,Romeo2012Springer}.

The theoretical analysis of nonlinearities and their effect on a topologically non-trivial interface is presented in~\cite{Pal2018PRE}, where results suggest lack of robustness of TP modes obtained through SIS breaking in the presence of a nonlinear interface. The present work sets the objective of observing this behavior experimentally. To this end, a nonlinear lattice consisting of units that interact through permanent magnets is modeled, assembled and then tested. Magnetic interactions provide the means for modulating the strength of the lattice coupling through proper adjustment of the interatomic spacing, and naturally introduce nonlinearities as the amplitude of wave motion increases. Specifically, topological protection is induced and subsequently verified via SIS breaking at a selected location, and is shown to undergo an interface-to-bulk transition for increasing amplitude. This occurs solely as a result of amplitude-dependent stiffness softening of the magnetic interaction, without requiring changes in the system's physical topology. 

Following this introduction (Sec.~\ref{Sec. Intro}), Sec. \ref{Sec.Theo} is devoted to the description of the considered lattice, its main physical parameters and the study of its corresponding analytical model, both in linear and nonlinear regimes. The experimental investigations are described in Sec.~\ref{Sec.Exp}. Finally, Sec.~\ref{Sec.Conclusions} summarizes the key findings of the study and highlights potential extensions. Three Appendices supplement the work.

\section{Lattice configuration and analytical results}\label{Sec.Theo}
\begin{figure}[ht] 
	\centering
	\subfigure[]{\includegraphics[width=0.4\textwidth]{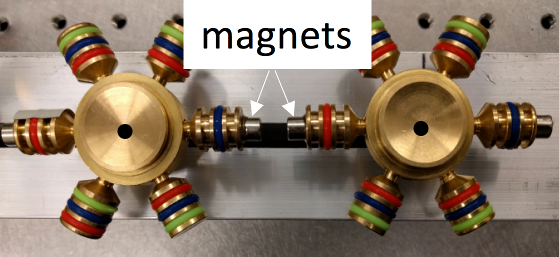}}\\
	\subfigure[]{\includegraphics[width=1\textwidth]{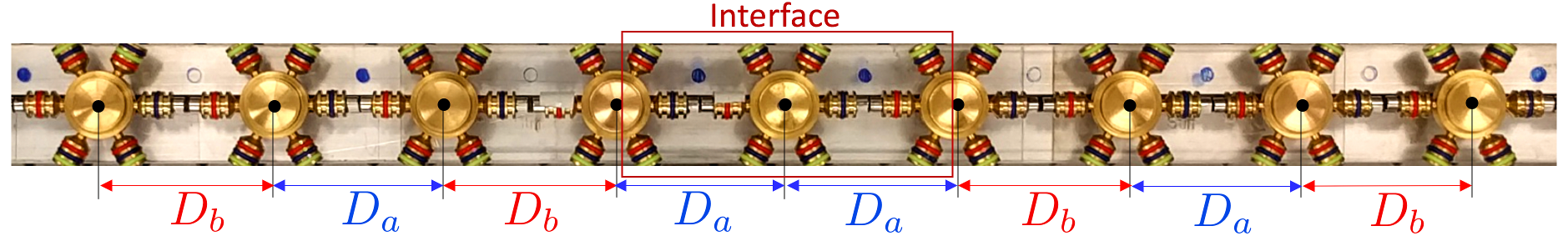}}
	\caption{ One-dimensional spinner lattice. (a) Detail of two interacting spinners, and (b) diatomic chain with interface generated through spatial inversion symmetry (SIS).}
	\label{Fig. Physical Chain}
\end{figure}

The investigations on TP and nonlinearities presented in~\cite{Pal2018PRE} have shown that localized modes arise at the interface between two spring-mass chains that are inverted copies of each other. In the presence of nonlinearities, amplitude-dependent frequency shifts cause the localized TP mode to migrate to the bulk spectrum. This behavior is further investigated in this paper through the physical implementation of a 1D lattice consisting of a dimer chain of spinners \cite{Prodan2018}, see Fig.~\ref{Fig. Physical Chain}. Each spinner is bolted to a linear guide, which fixes its position while letting it free to rotate about an axis perpendicular to the page. The spinners are coupled through permanent magnets in attraction that provide a force that tends to maintain the spinners in the aligned position (Fig.~\ref{Fig. Physical Chain}(a)). The magnitude of magnetic interactions is strongly related to the distance between the magnets, which is defined by the spacing between the spinners. Such spacing is here modulated to implement a dimer lattice configuration whereby the interaction coefficients are defined by two distance values, namely $D_a$ and $D_b$ (Fig.~\ref{Fig. Physical Chain}). An interface is created by joining the lattice with its mirror copy at a defined location as a result of broken SIS (Fig.~\ref{Fig. Physical Chain}(b)).

\subsection{Analytical model}\label{Sec.Geo}
\begin{figure}[ht] 
	\centering
	\includegraphics[width=1\textwidth]{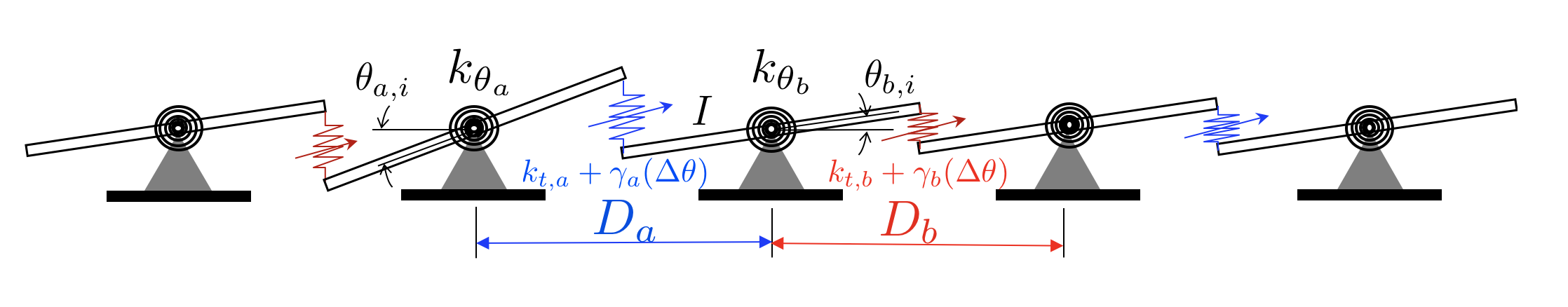}
	\caption{Schematic of analytical model with key physical parameters.}
	\label{Fig. Model Schematic}
\end{figure}

A simplified model is formulated according to the configuration of Fig.~\ref{Fig. Model Schematic}. The dynamic behavior of each spinner is described by its rotation angle $\theta$, and governed by the spinner inertia $I$ and by the interaction with its neighbors. Such interaction is evaluated based on the model of the magnetic force exchanged by the permanent magnets mounted on the spinner's pegs, which can be approximated to varying orders in terms of the angular positions of the spinners. Details of the evaluation of the magnetic interactions and their simplified description can be found in Appendix~\ref{Sec.AnalogMagnetic}.

According to the approximations made and the derivations reported in the Appendix~\ref{Sec.AnalogMagnetic}, the equations of motion for the $i$-th unit cell can be expressed as follows:

\begin{widetext}
\begin{equation}\label{Eq.Gov}
\begin{split}
I \ddot{\theta}_{a,i}+k_{\theta}  \theta_{a,i} + k_{t,a}(\theta_{b,i}+\theta_{a,i})+k_{t,b} (\theta_{a,i}+\theta_{b,i-1})+\gamma_a(\theta_{b,i}+\theta_{a,i})^3+\gamma_b(\theta_{a,i}+\theta_{b,i-1})^3 & =  0\\
I \ddot{\theta}_{b,i}+k_{\theta}  \theta_{b,i}+k_{t,b} (\theta_{a,i+1}+\theta_{b,i})+k_{t,a}(\theta_{b,i}+\theta_{a,i})+\gamma_b(\theta_{a,i+1}+\theta_{b,i})^3+\gamma_a(\theta_{b,i}+\theta_{a,i})^3 & = 0
\end{split}
\end{equation}
\end{widetext}
where $I$ is the inertia of each spinner, $k_{\theta_a}, k_{\theta_b}$, $k_{t_a}, k_{t_b}$ are the linear interaction coefficients, while $\gamma_a, \gamma_b$ define the nonlinear interaction coefficients.  The equations for the inverted unit cell are formally identical, with the proper switching of the subscripts, and are reported in Appendix~\ref{Sec.AnalogMagnetic} for brevity.

Analysis of the equations reveals that the motion of each spinner is governed by its rotary inertia, and by the magnetic interactions that in the linear regime manifest themselves as a term that is proportional to the rotation of each individual spinner. This effectively produces the effect of a torsional spring connected to the ground. An additional term couples neighboring spinners through a torque that is approximately proportional to the relative displacement between neighboring magnets in the direction transverse to the spinners chain, here measured by the sum of their respective rotation angles.
	
\subsection{Linear dispersion analysis and associated topology} \label{Sec.LinearTheory}
We first investigate the underlying linear behavior of the lattice, by considering small angular perturbations and neglecting the nonlinear terms in Eq.~\eqref{Eq.Gov}. We evaluate the dispersion properties for the infinite lattice by imposing a plane wave solution in the form $\theta_{p,i}=\theta_{p,0} e^{\rm{j} ( i \mu -\omega t)}$, where $i$ is an integer defining the location of the unit cell, $p=a,b$, $\rm{j}=\sqrt{-1}$, while $\omega$ denotes the angular frequency and $\mu$ the dimensionless wavenumber. Substituting these expressions in Eqs.~\eqref{Eq.Gov}, we obtain an eigenvalue problem that identifies the following two dispersion branches:
\begin{equation}
\begin{split}
\label{eq: linear dispersion}
\omega^2 & =\frac{1}{I} \left(k_{\theta}+k_{t,a}+k_{t,b}\right)\\
				   & \pm \frac{1}{I} \sqrt{k_{t,a}^2+k_{t,b}^2+2 k_{t,a} k_{t,b} \cos  \mu },
\end{split}
\end{equation}

\begin{figure}[hbtp] 
	\centering
	\subfigure[]{\includegraphics[width=0.45\textwidth]{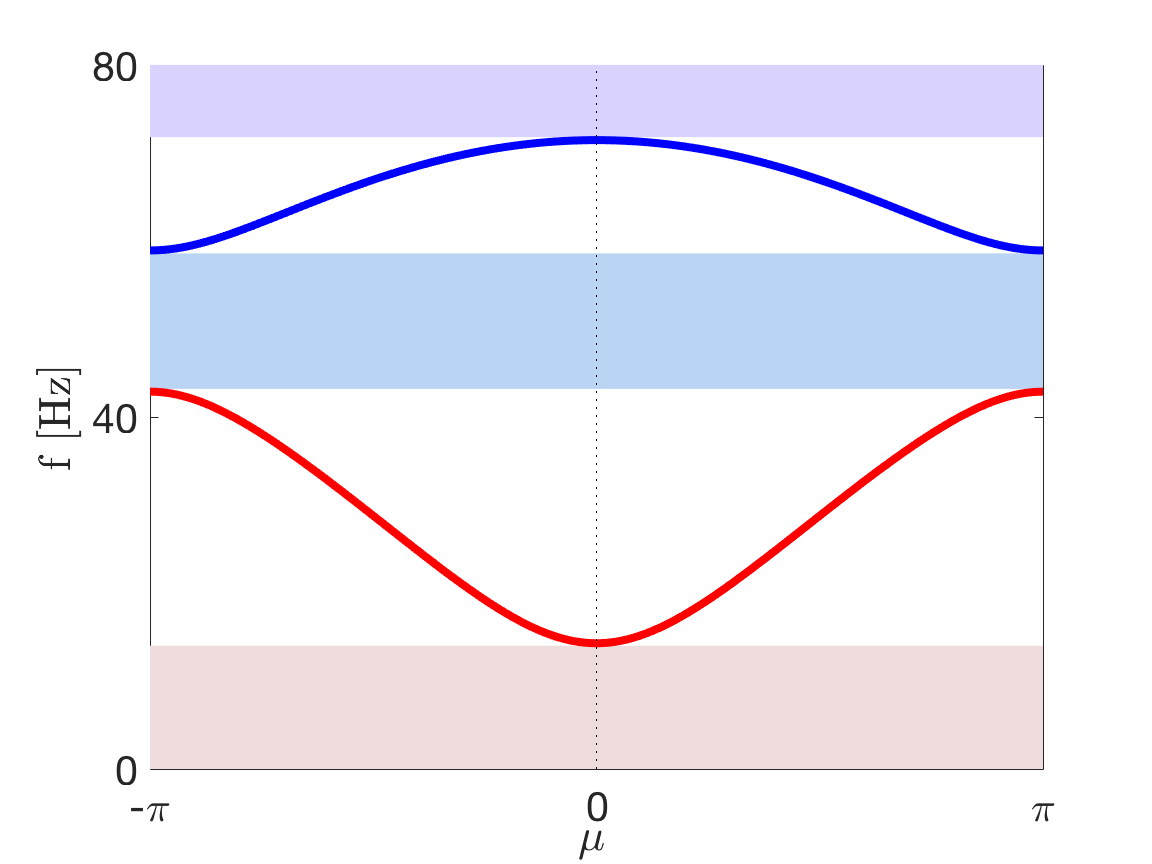}}
	\subfigure[]{\includegraphics[width=0.45\textwidth]{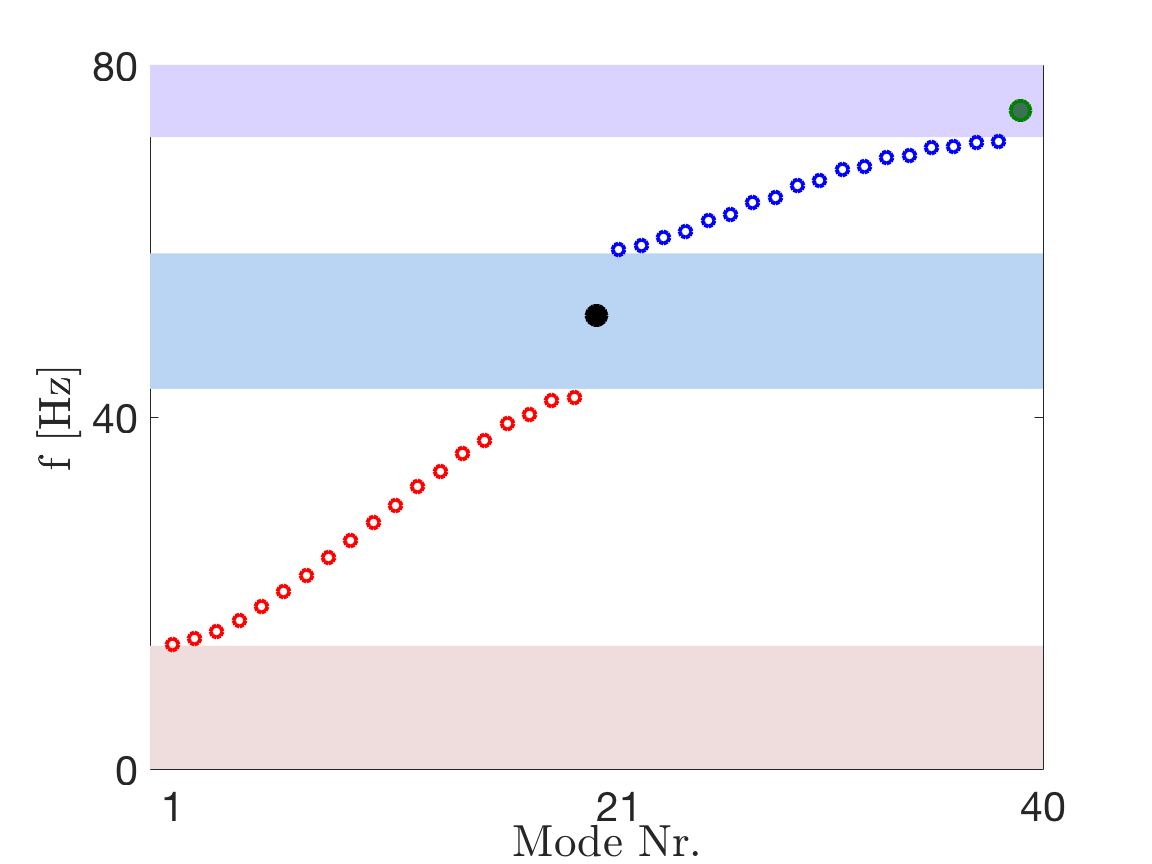}}\\
	\subfigure[]{\includegraphics[width=0.6\textwidth]{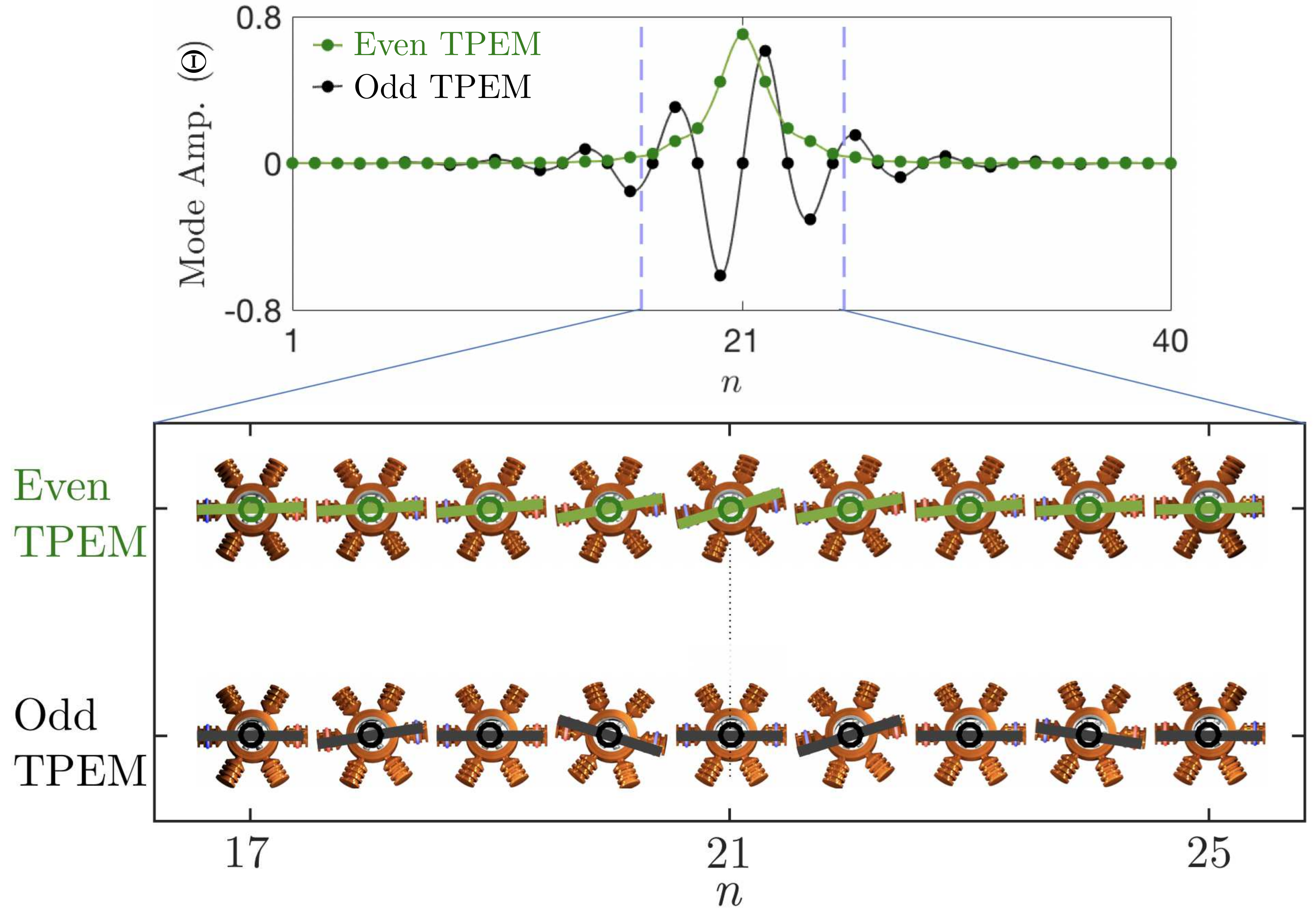}}
	\caption{(a) Linear dispersion diagram for the periodic lattices. (b) Eigenvalues for a 20+20 spinners lattice with the non-trivial interface showing the existence of two TP modes populating the second and third bandgap (black and green solid dots). Notice that there is only 39 eigenvalues because the first equation is removed since motion in spinner $n=1$ is imposed. (c) Corresponding eigenvectors illustrating the symmetric (even) and antisymmetric (odd) spatial distribution of the TP modes.}
	\label{Fig.DD analytical}
\end{figure}

This lattice features two dispersion branches separated by three bandgaps (Fig.~\ref{Fig.DD analytical}(a)). The first bandgap starts at zero frequency and extends up to a frequency cut-off at $\mu=0$, which is the result of the grounding constants $k_{\theta}$. Breaking of spatial inversion symmetry by inverting the order of the distance modulations, produces dispersion curves that differ in terms of the associated topological invariants. Specifically, the topological properties of the second and third bandgaps can be switched by permutation of the intra-cell and inter-cell connecting springs, i.e. inverting the unit cell, or by considering $k_{t,a}>k_{t,b}$ or vice versa, i.e. $k_{t,b}>k_{t,a}$. The topological invariant, the Zak phase \cite{Zak1985} in the case of a 1D lattice, is evaluated through numerical integration of the eigenvector change along each band as described in~\cite{Pal2018PRE,zhou2017optical,xiao2015geometric}. It is found that the Zak phase is $\mathcal{Z}=\pi$ for both dispersion bands when $k_{t,a}<k_{t,b}$, while it is $\mathcal{Z}=0$ otherwise. Hence, the interface of Fig.\ref{Fig. Physical Chain}(b) connects two lattices with same bandgaps, but inverted geometry and different bands topology.  Thus, the interface supports TP modes whose frequency can be predicted from the solution of the eigenvalue problem for a finite system. The eigenvalues obtained for two inversed lattices with 20 spinners each confirm the existence of 3 bandgaps, along with the presence of two TP modes inside the second and third gap (black and green solid dots, respectively). The expected localized nature of the modes at the interface is illustrated by the eigenvectors corresponding to the two TP frequencies (Fig.~\ref{Fig.DD analytical}(c)), which show the limited penetration of each mode in the bulk, and illustrate the modes' distinct spatial profiles, whereby the lower frequency mode is odd relative to the interface, while the higher frequency mode is symmetric, or even, with respect to it.
 
\subsection{Effects of nonlinear interactions}\label{Sec.NonLinearTheory}
Next, we evaluate the effect of increasing amplitude on the eigenvalues and associated eigenmodes of the system. To this end, we consider the governing equations for the finite $N+N=40$ system with interface, which are obtained from the assembly of equations in Eq.~\eqref{Eq.Gov}. Assuming harmonic motion $\theta_n=\Theta_n e^{j\omega t}$ and applying harmonic balance, we obtain the general matrix form:
\begin{equation}\label{Eq. nonlinear gov equation}
\bK(\bTheta)\bTheta= \omega^2 I \bTheta.
\end{equation}
where $\bm{\Theta}=[\Theta_{a,1}, \Theta_{b,1},...,\Theta_{a,N}, \Theta_{b,N}]^T$ is a vector including the complex amplitudes of all angular degrees of freedom of the lattice, $\bK(\bTheta)$ denotes the effective stiffness matrix and $\btheta=\bTheta e^{j\omega t}$. For low amplitudes $|\bTheta|\ll 1$, the stiffness matrix $\bK$ is independent of $\bTheta$ and the solution is straightforward. However, when nonlinearities play a role the effective stiffness matrix depends on the amplitudes of motion, which requires an iterative analysis. Specifically, we use a Newton-Raphson scheme~\cite{AZRAR1999}.

To write the nonlinear governing equations in canonical form, Eq. \eqref{Eq. nonlinear gov equation} is rearranged as:
\begin{equation}\label{Eq. nonlinear gov equation2}
\left[ \bK(\bTheta)- \omega^2 I \right] \bTheta = 0.
\end{equation}
This system of $2N$ equations has $2N+1$ unknown variables $\{\bTheta,\omega\}$, and therefore infinite solutions. To extract specific $\{\bTheta, \omega\}$ pairs, we impose particular values to the total amplitude of the chain $A$, defined as the $L_2$ norm of $\bTheta$. Thus we add the additional equation $|\bTheta|_2-A=0$, where $A$ has a numeric value. When $A\rightarrow 0$ is imposed, the linear solution is recovered.

We start by solving for a small value of $A$ (e.g. $A=10^{-3}$), and we use the linear eigenvector-eigenvalue pair $\{\bTheta_l,\omega_l\}$ as initial guess. The linear eigenvector $\bTheta_{l}$ is simply scaled as $\bTheta_{g}=\bTheta_{l}/|\bTheta_{l}|_2A$ and the linear eigenvalue $\omega_l$ is used as is. This way we ensure that the initial guess $\bTheta_{g}$ is the eigenvector of the linear problem and that its total amplitude $|\bTheta_{g}|$ is $A$. The algorithm yields a new solution that is then used as the initial guess for a slightly higher value of $A$, and so on. With this procedure we calculate the evolution of the eigenvalue-eigenvector pair for increasing values of total amplitude $A$.
 
\begin{figure}[hbtp] 
	\centering
	\subfigure[]{\includegraphics[width=0.45\textwidth]{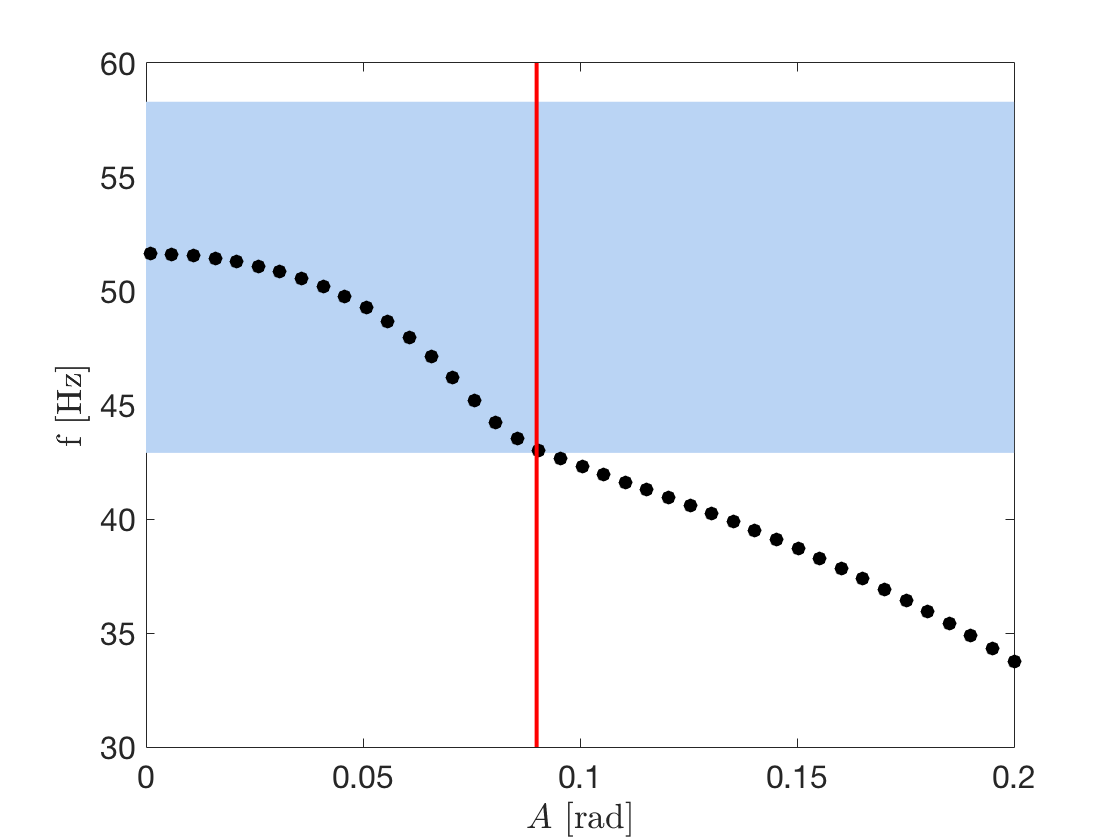}}
	\subfigure[]{\includegraphics[width=0.45\textwidth]{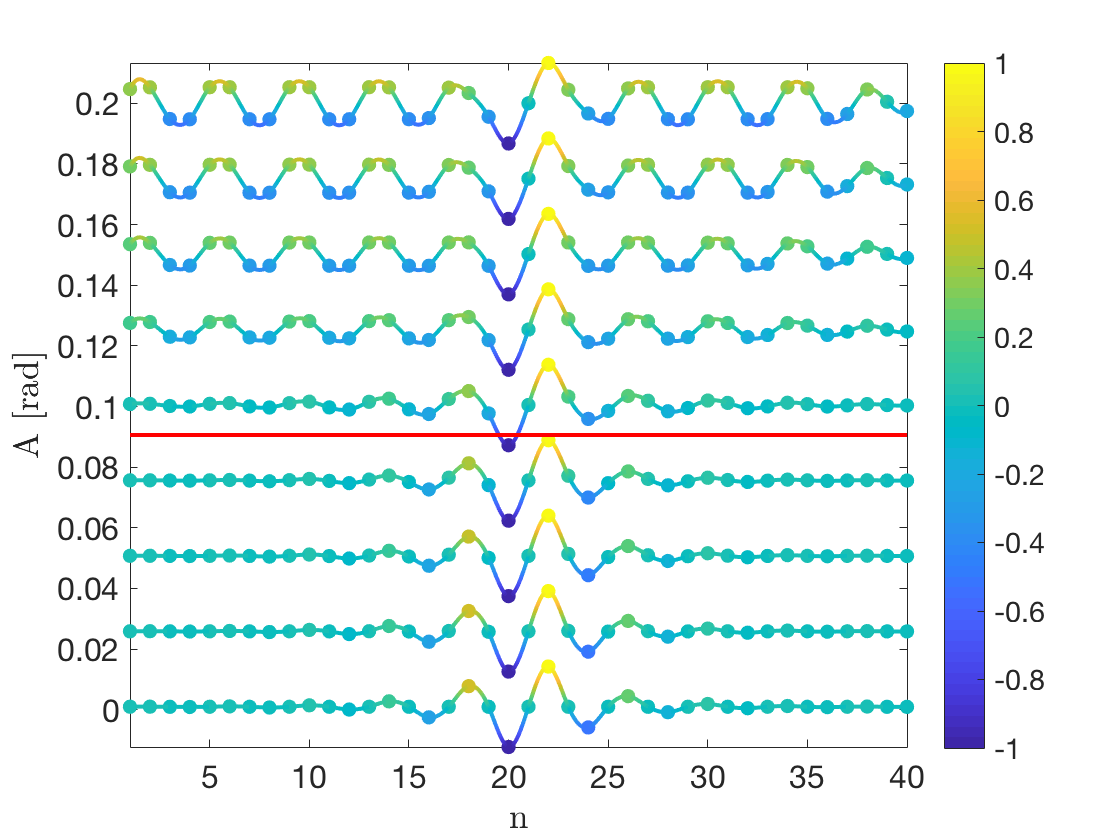}}
	\subfigure[]{\includegraphics[width=0.45\textwidth]{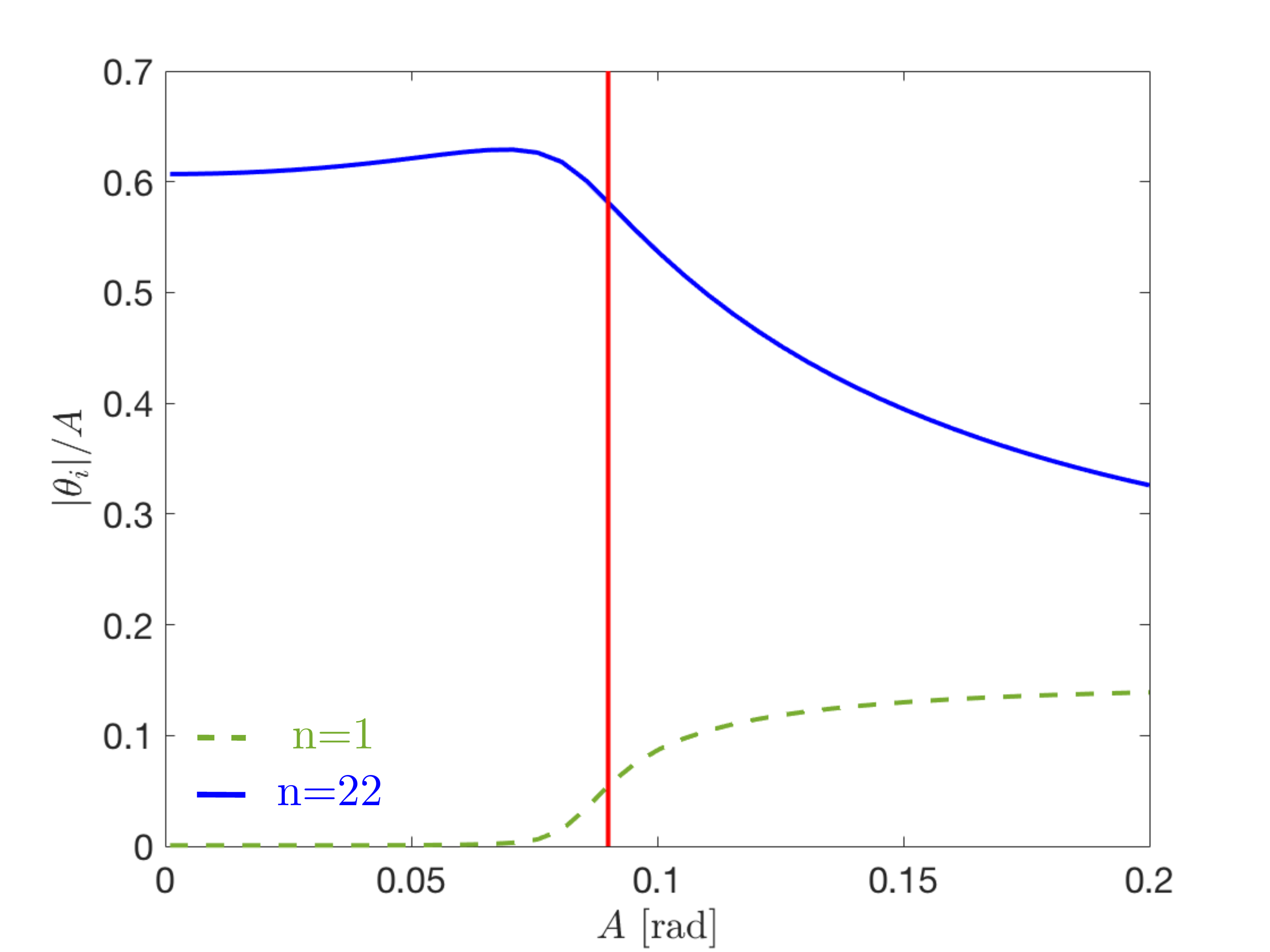}}
	\caption{Effects of nonlinearities on the odd TP mode. (a) Variation of the eigenvalue versus amplitude (black dots); shaded blue area outlines the linear bandgap, while the vertical solid red line marks the amplitude corresponding to the interface-to-bulk transition at $A\approx0.09$ rad. (b) Variation of eigenmodes in terms of amplitude (colorbar is associated to the normalized magnitude of each mode). (c) Variation of normalized magnitudes at locations $n=1$ (dashed green line) and $n=22$ (thick blue line) and transition amplitude (solid red line).}
	\label{Fig.Nonlinear.analytic}
\end{figure}

\color{black}
Depicted in Figs.~\ref{Fig.Nonlinear.analytic}(a)-(d) are results for the odd mode for the values of $\gamma_{a(b)}=-366(-188)$ Nm/rad$^3$ (see Appendix~\ref{Sec linear coefficients}). Results show that the nonlinear ``eigenfrequency" decreases with amplitude, along with an amplitude-dependent transition whereby the frequency exits the bandgap (shaded blue area in Fig.~\ref{Fig.Nonlinear.analytic}.a) and enters the bulk spectrum of the linear system. This is consistent with the negative value of $\gamma_{a(b)}$ that defines a softening nonlinearity in the connecting springs, by which their effective stiffness decreases for increasing total amplitude $A$. When the nonlinear eigenvalue abandons the bandgap, the bulk attenuation of this otherwise localized wave mode no longer holds, and the wave mode extends to the bulk. This is illustrated in Fig.~\ref{Fig.Nonlinear.analytic}(b), which presents the variation of the corresponding eigenvectors for increasing amplitude $A$. In the figure, the colors are associated with the magnitude of each mode normalized to its maximum value, i.e. $\bTheta(A)/|\bTheta(A)|_\infty$. Also, the markers correspond to the normalized angular motion of the individual spinner, while the continuous solid lines are spline interpolations for improved visualization. Both plots in Fig.~\ref{Fig.Nonlinear.analytic}(a),(b) illustrate the occurrence of an interface-to-bulk transition as the amplitude of wave motion increases, and show the importance of nonlinearities. The transition is denoted by the thick, solid red lines in both figures at $A\approx0.09$ rad and is further illustrated in Fig.~\ref{Fig.Nonlinear.analytic}(c), which compares the magnitude of the eigenvector at spinner $n=22$ close to the interface (solid blue line), and away from the interface at $n=1$ (dashed green line). For low amplitudes, motion at $n=1$ is very limited, and negligible compared to the motion at the interface. As amplitude increases, there is an evident increase in motion at the beginning of the chain ($n=1$) as a result of the mode becoming global in nature and no longer localized at the interface. A thick red line at $A\approx0.09$ rad is added to the plot for reference purposes.

\section{Experiments} \label{Sec.Exp}
\begin{figure}[hbtp]
	\centering
	\includegraphics[width=0.99\textwidth]{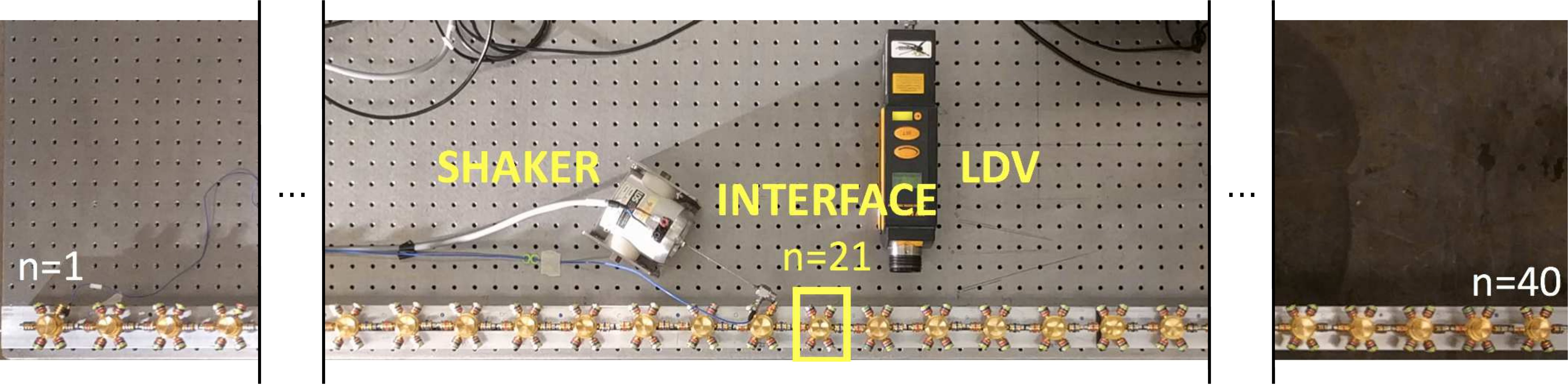}
	\caption{Physical 40 spinner system mounted on a beam. Distances $D_a=7\textrm{ mm},\,D_b=6\textrm{ mm}$ are denoted by an empty and a full blue circle respectively (rubber band colors indicate magnets polarity). This is the setup when motion is imposed to spinner 20.}
	\label{Fig.Setup}
\end{figure}

We experimentally evaluate the existence of TP modes and the influence of amplitude and associated nonlinearities through the 40 spinner array shown in Fig.~\ref{Fig.Setup}. The spinners are bolted to a longitudinal aluminum beam at distances $D_a$ and $D_b$. The magnets employed are bonded to the pegs of the spinners, with aligned magnetization vectors poled in attraction. The method used to experimentally characterize the magnetic interaction as a function of the distance between the magnets is described in Appendix~\ref{Sec coefficients}. The key relevant model parameters identified through those experiments are listed in Table \ref{Table.ExpValC1C2}. Additional details of geometric properties of the magnets, spinners and the chain are provided in Appendix~\ref{Sec experimental set up}.

In the experiments, excitation is provided by an electrodynamic shaker controlled by a signal generator that provides the desired input. Specifically the signals used in the experiments are a white noise signal band-limited to the frequency range of interest ($0-80$ Hz) and a sine wave at the target frequency and amplitude. The response of the spinner array is recorded by a single point Laser doppler vibrometer (LDV) pointed at selected locations. Experiments are conducted for excitation applied at spinner $n=1$ at the left boundary of the array, and at spinner $n=20$ close to the interface. The first configuration evaluates the transmissibility through the array, while the excitation right at the interface (n=20) directly probes the TP modes and investigates changes as a function of amplitude. The configuration with excitation at spinner $n=20$ is shown in Fig.~\ref{Fig.Setup}. Video recordings of the response of the spinner arrays are also taken through a high speed camera, the results of which are processed to provide the spatial distribution of the response and show mode localization and to produce the animations presented as part of the Supplementary Material (SM).

In the SM videos, we show the spinners chain oscillating at the nonlinear normal frequencies of three different values of the amplitude denoted as low $A=0.002$ rad, medium $A=0.070$ rad and high $A=0.179$ rad. We superimpose a circle on top of every spinner whose radius is proportional to the spinner amplitude of motion $|\Theta_n|$ for improved visualization. The interiors of these circles are colored to indicate the instantaneous phase of each spinner measured as the argument of the complex number $\Theta_n e^{j\omega t}$ in absolute value, going from cyan in the lowest value of the spinner oscillation $|\arg{(\Theta_n e^{j\omega t})}|=\pi$ to magenta in the highest one $|\arg{(\Theta_n e^{j\omega t})}|=0$. A small oscillating white circle is also attached to the perimeter of each circle to this end.
	
For verification of the LDV measurements, one point of each spinner, located next to the one of the magnets, is tracked to extract the spinner motion $\theta_n$ from the videos. The points are marked in the animations with a blue dot surrounded by a red square. We track the motion by comparing the relative position of the pixel set inside the red square among subsequent frames.

\subsection{Low amplitude response: linear behavior}\label{Sec.ExpLinear}
\begin{figure}[hbtp]
	\centering
	\includegraphics[width=0.60\textwidth]{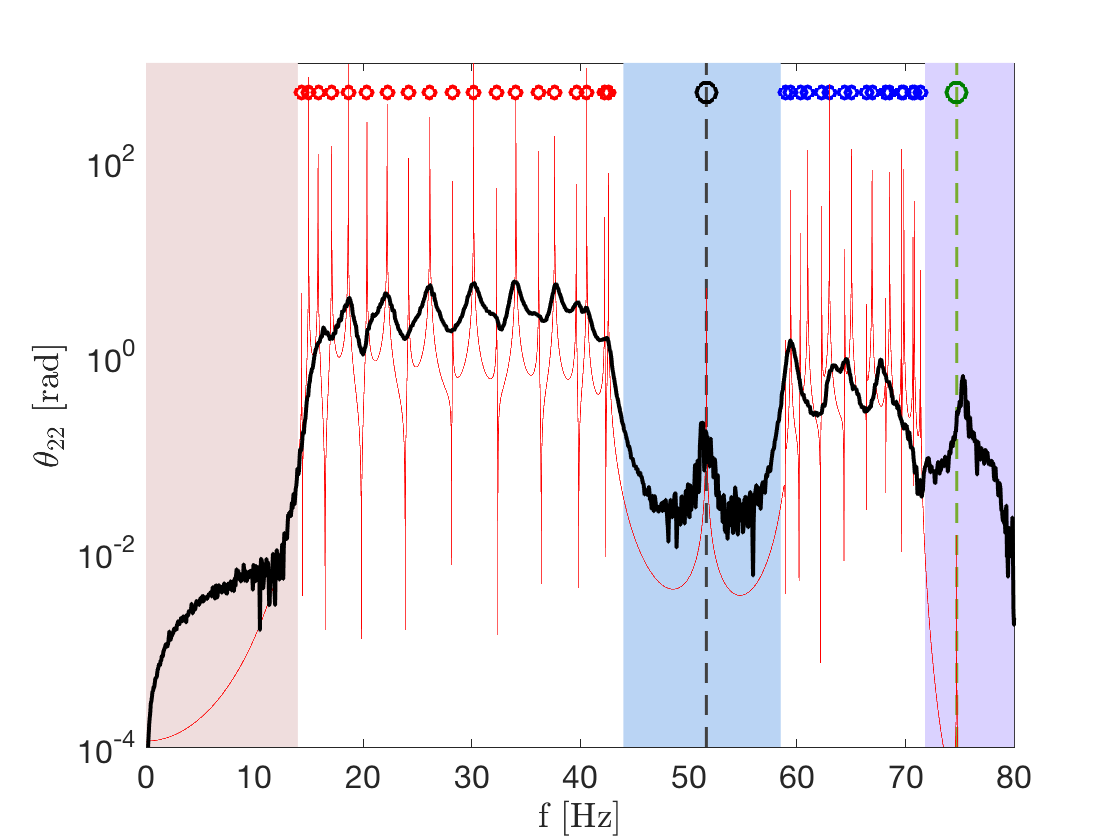}
	\caption{Experimental frequency response at spinner $n=22$ in the \textit{linear regime} for white noise excitation at $n=20$. For reference, the theoretical predictions are reported in the thin red line, along with the theoretical eigenvalues (red and blue circles) and the frequency corresponding to the TP modes (black and green circles and vertical dashed lines). The shaded beige, cyan and purple regions denote the analytical linear bandgaps.}
	\label{Fig.Results.Linear}
\end{figure}

As in the analytical investigations, we first probe the linear behavior of the system by evaluating its dynamic behavior at low amplitude. To this end, we measure the frequency response at $n=22$, which immediately follows the interface, for white noise excitation applied at spinner $n=1$ during 20 seconds, and averaged for 150 repetitions. The results are presented in Fig.~\ref{Fig.Results.Linear} (black solid line). For reference the figure also reports the corresponding analytical predictions (red solid line), along with the predicted eigenvalues (red circles), and the frequency bandgaps (shaded beige, cyan and purple regions). The results show a good match  between analytical and experimental results, and confirm the overall behavior of the system, including the existence of bandgaps and of the two TP modes, also highlighted in the figure. 

\subsection{Amplitude dependent response: nonlinear regime}\label{Sec.ExpNonlinear}
Finally we investigate amplitude effects around the frequency of the odd TP mode. Therefore, we impose harmonic motion at spinner $n=20$ and record the its acceleration, along with the applied force, which is measured by a load cell mounted on the stinger connected to the shaker, and the velocity of spinner $n=22$. All results presented herein are at steady-state, as a result of experiments conducted for frequency varying between 35 Hz and 55 Hz, and amplitude of imposed motion $\theta_{20}=\Theta_{20} e^{j\omega t}$ increasing approximately between $|\Theta_{20}|=0.001$ rad and $|\Theta_{20}|=0.07$ rad. Since the shaker is controlled in open-loop, we control the amplitude of the electronic signal that excites it, and $\Theta_{20}$ is evaluated as the first harmonic of the motion of spinner $n=20$, recorded by an accelerometer. The amplitude $\Theta_{22}$ of spinner $n=22$ is also calculated as the first harmonic of its motion $\theta_{22}$, measured with the LDV. The amplitude of applied force $f_0$ is calculated as the first harmonic of the instantaneous force measured by the load cell. Second and higher harmonics of all the measurements have been found more than an order of magnitude lower than the first harmonic.

\begin{figure}[hbtp]
	\centering
	\subfigure[]{\includegraphics[width=0.45\textwidth]{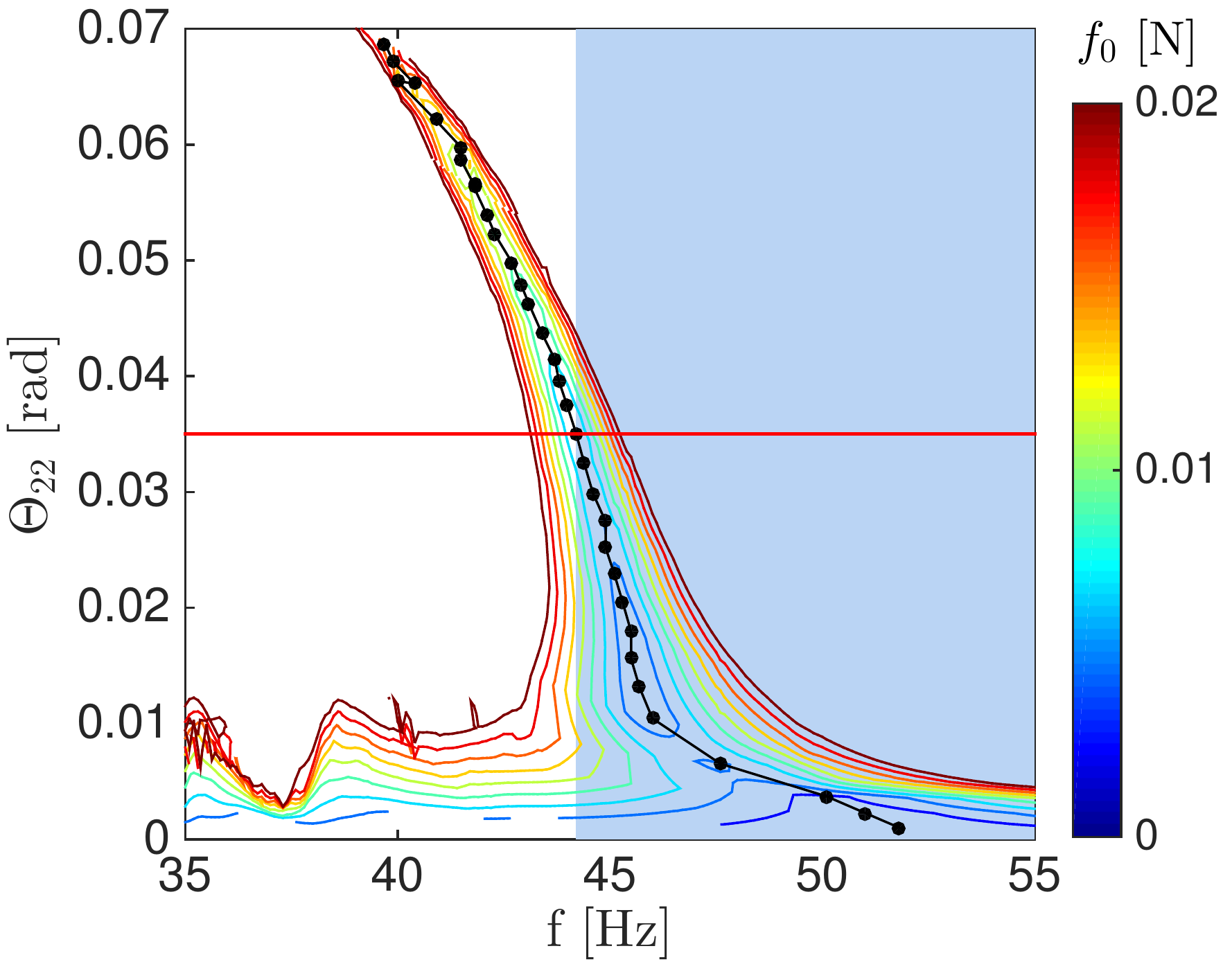}}\\
	\subfigure[]{\includegraphics[width=0.45\textwidth]{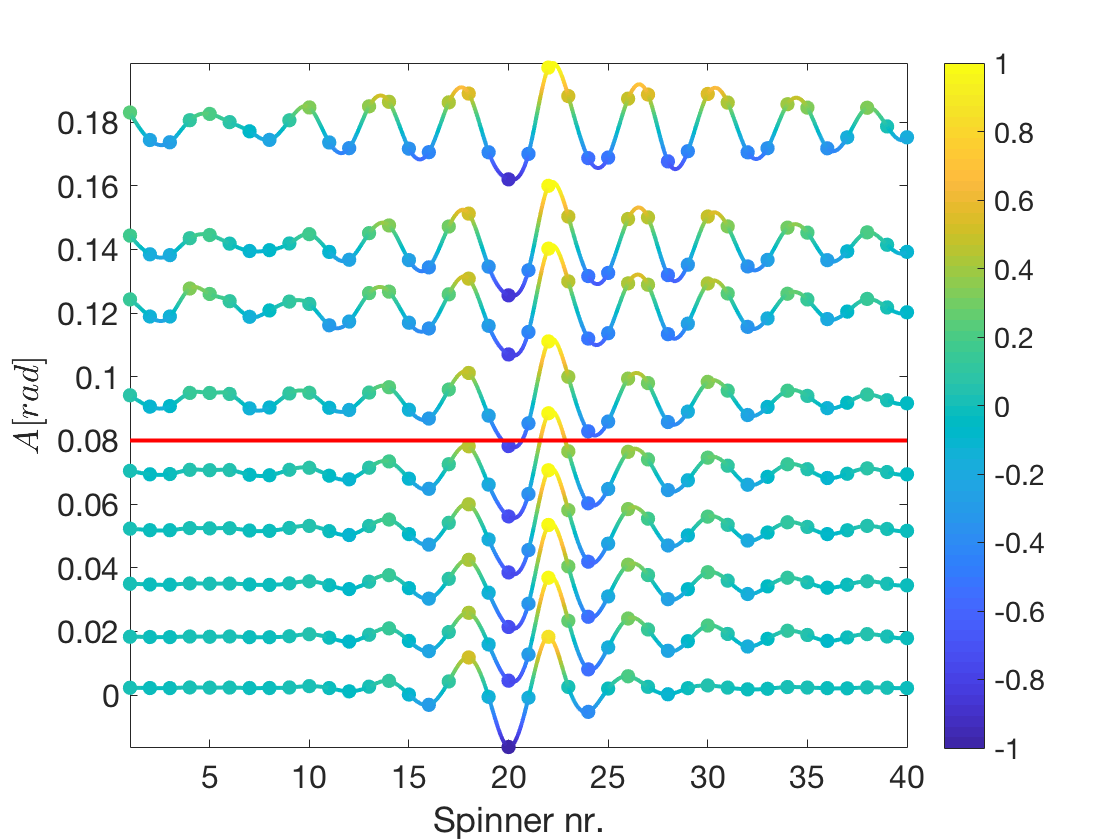}}
	\subfigure[]{\includegraphics[width=0.45\textwidth]{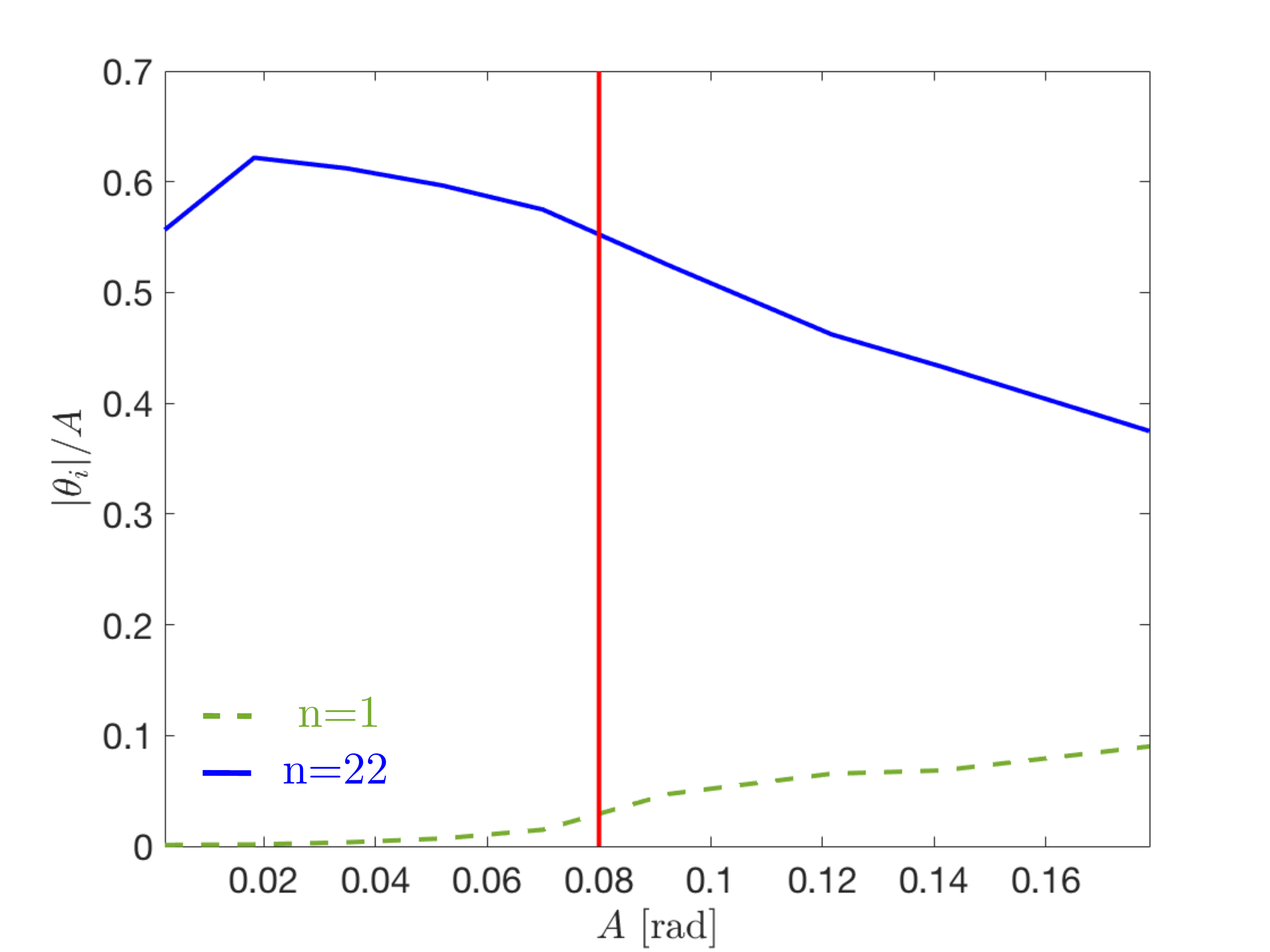}}
	\caption{Experimentally observed effects of nonlinearities on the odd TP mode. (a) Amplitude $|\Theta_{22}|$ versus frequency relation for nonlinear normal modes. Shaded blue area outlines the linear bandgap, while the horizontal solid red line marks the amplitude corresponding to the interface-to-bulk transition at $|\Theta_{22}|\approx0.035$ rad. The contours represent the frequency-response correlation for oscillations excited at constant force amplitude. (b) Variation of steady-state dynamic deformed shapes in terms of total amplitude $A$ (the colorbar is associated to the normalized magnitude of each mode). The transition occurs at amplitude $A\approx0.08$ rad. (c) Variation of normalized magnitudes at locations $n=1$ (dashed green line) and $n=22$ (thick blue line) and transition amplitude (solid red line).}
	\label{Fig nonlinear exp}
\end{figure}

Each experiment produces a triplet of values: the amplitude of the response $\Theta_{22}$, its frequency, and the amplitude of the applied force $f_0$. Mapping these values through a series of experiments leads to a surface that correlates frequency, amplitude of response and amplitude of applied force. The surface can be represented as contours that relate frequency and amplitude of response at constant applied force. In this representation, resonance frequencies are identified as points of minimum required force, i.e. as the valley of these surface. The results are presented in Fig.~\ref{Fig nonlinear exp}(a), where the natural frequencies are represented by the black dotted line in the figure, which forms a typical backbone curve. The backbone curve presents a sharp change in slope as the frequency leaves the bandgap (shaded blue region), which presumably indicates a transition in dynamic behavior. In addition, we record the dynamic deformed shape for excitation at the backbone frequencies. The measurements are conducted by repeating LDV measurements at each spinner location and then combining the corresponding amplitude and phase to obtain each of the curves shown in Fig.~\ref{Fig nonlinear exp}(b). For these, the LDV head is manually moved between locations and the data acquisition device is programed to synchronize the measurements by starting them always at the same time interval after the excitation signal is triggered. The figure presents the change in the dynamic deformed shapes as a function of total amplitude $A=|\bTheta|_2$, which clearly illustrates how the lattice exhibits the predicted change in the linear-regime TP mode, and documents its transition from being localized at small amplitudes, to bulk mode for higher values of $A$. As in the analytical results, the amplitude of motion at spinner $n=1$ is negligible in the linear regime, but grows for increasing nonlinearities (Fig. \ref{Fig nonlinear exp}(c)). Evidence of a transition, although not as sharp as the one predicted by the theoretical model (in Fig. \ref{Fig.Nonlinear.analytic}), is marked by the vertical solid red line at $A=0.08$ rad.

An alternative visualization of the transition is obtained by recording the motion of the spinners through a high speed camera. The experiments are conducted by repeating the measurements over 15 separate portions of the lattice, as the entire length exceeds the aperture of the camera. Upon the recording, the measurements are then phase-matched and stitched to obtain a single recording for an assigned amplitude of motion. Snapshots of the deformed configurations of the chain for 3 values of amplitude $A$ are shown in Fig.~\ref{Fig snapshots}. As the angular rotation of the spinners in all cases remain relatively hard to observe from the pictures, circles of radius proportional to the amplitude of motion are superimposed to each spinner to facilitate visualization and to better appreciate the extent of the penetration of the mode into the bulk. Such penetration is very limited for low amplitudes Fig.~\ref{Fig snapshots}(a), as the mode is strongly localized at the interface, and progressively increases for higher values of amplitude to eventually reach the end of the chain in the case shown in Fig.~\ref{Fig snapshots}(c). Also for visualization purposes, the interior of the circles indicating amplitude is colored to indicate the instantaneous phase of each spinner measured as the argument of the complex number $\Theta_n e^{j\omega t}$, going from cyan in the lowest value of the spinner oscillation $\arg{(\Theta_n e^{j\omega t})}=\pi$ to magenta in the highest one $\arg{(\Theta_n e^{j\omega t})}=0$. To this end, a small white circle is added to the perimeter of the circles.

\section{Conclusions}\label{Sec.Conclusions}
The paper investigates the occurrence of topologically protected interface modes produced by broken spatial inversion symmetry. Experimental observations are conducted on a one dimensional dimer chain consisting of spinners coupled through permanent magnets. Spatial modulation of the interaction strength relies on setting the distance between magnets of neighboring spinners. Guided by a simplified analytical model, dynamic measurements highlight the presence of frequency bandgaps and of topologically protected interface modes whose frequencies lie inside the gaps. The experiments also probe the behavior of the chain when nonlinearities affect lattice interactions. A softening-type nonlinearity cause the frequency of the topologically protected modes to progressively merge with the linear bulk bands, causing an interface-to-bulk transition of the corresponding mode. Such transition is first predicted by the analytical model, and then confirmed by the measured response of the chain. Laser vibrometry and full field optical capture of the dynamic deformed configurations of the lattice are employed to quantify and characterize the interface localization of the topologically protected modes, and their extinction as the amplitude of motion increases. A transition amplitude is predicted numerically and also observed experimentally, with a good level of agreement. The study paves a path towards the understanding of the robustness of topologically protected modes and lack thereof in the presence of the type of nonlinearities investigated as part of this study. The results also suggest a potential mechanism for the control of localization and the transition to bulk propagation that exploits topological protection in conjunction with nonlinear interactions.

\section*{Acknowledgments}
The work is supported by the Army Research Office through grant W911NF-18-1-0036, and by the National Science Foundation through the EFRI 1741685 grant. The authors want to acknowledge Camille and Emil Prodan for sharing the idea of employing spinners as part of an effective experimental framework.

\bibliography{Literature_MagnetSpinners}
\bibliographystyle{apsrev4-1}

%\onecolumngrid
%\input{Figures}
%%%%%%%%%%%%%%%%%%%%%%%%%%%%%%%%%%%%%%%

 \clearpage

\begin{figure}[hbtp]
	\centering
	\subfigure[]{\includegraphics[width=0.9\textwidth]{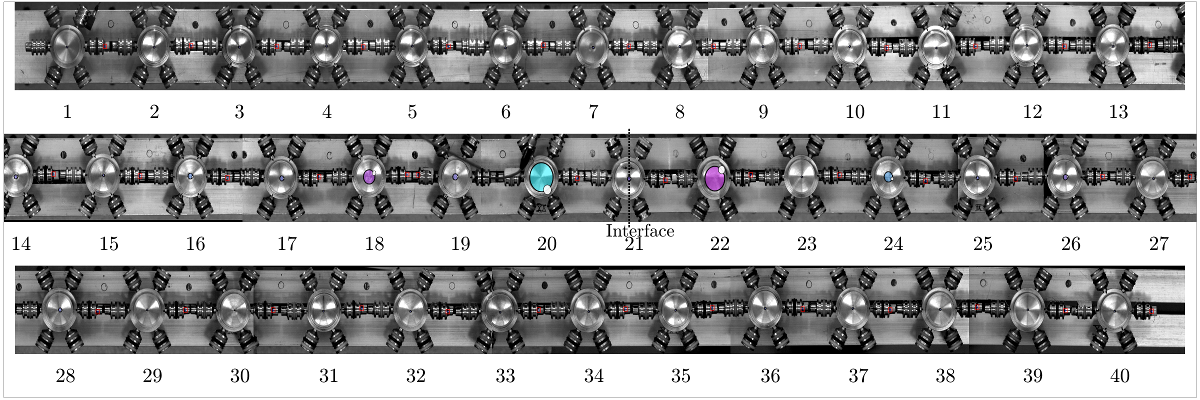}}\\
	\subfigure[]{\includegraphics[width=0.9\textwidth]{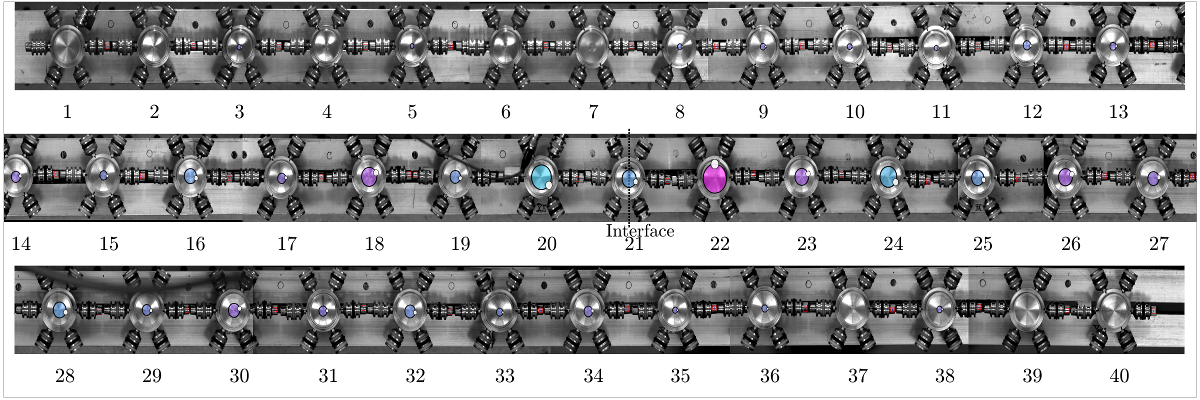}}\\
	\subfigure[]{\includegraphics[width=0.9\textwidth]{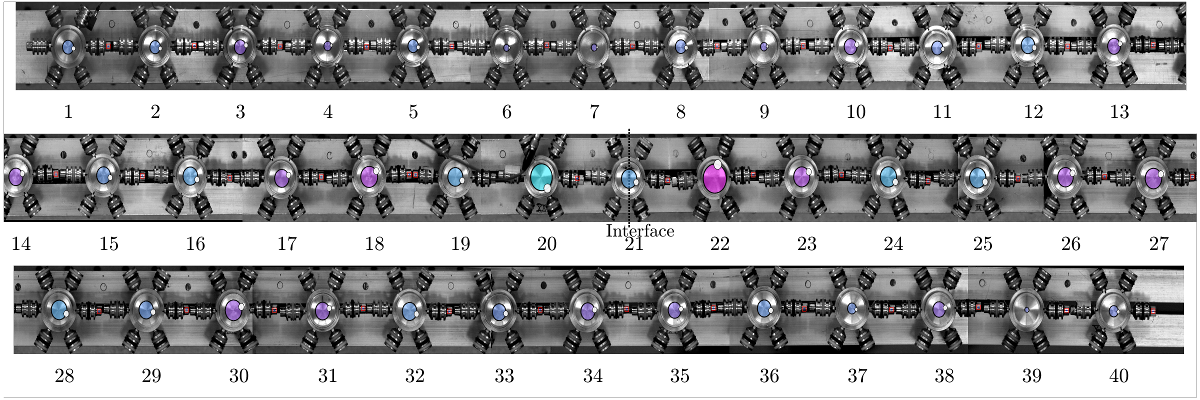}}
	\caption{Experimentally measured snapshots of the chain motion for increasing values of amplitude: (a) low amplitude $A=0.002$ rad; (b) medium amplitude $A=0.070$ rad, and (c) high amplitude $A=0.179$ rad. Circles of radius proportional to the normalized angular motion of each spinner are superimposed to the picture to aid visualization.}
	\label{Fig snapshots}
\end{figure}
%%%%%%%%%%%%%%%%%%%%%%%%%%%%%%%%%%%%%%%
%APPENDIX

\clearpage

\appendix
\numberwithin{equation}{section}
\numberwithin{figure}{section}
\numberwithin{table}{section}

\section{Model of magnetic interaction}\label{Sec.AnalogMagnetic}
The magnetic force is evaluated by computing the interaction between magnetically rigid dipole moments  $\bmm_a$ and $\bmm_b$, which is given by~\cite{Petruska2013}: 
\begin{equation}\label{Eq.MagDipoleForce}
%\begin{split}
\bm f_{ba}  =-\frac{3 \mu_0}{4 \pi d^5} (  {\bd} \left(  {\bmm}_a\cdot  {\bmm}_b\right)  +  {\bmm}_a \left(  {\bd}\cdot {\bmm}_b\right)  
+  {\bmm}_b ( {\bd}\cdot {\bmm}_a) - \frac{5  {\bd}}{d^2} \left(  {\bd}\cdot  {\bmm}_a\right)  \left(  {\bd} \cdot {\bmm}_b\right) ) 
%\end{split}
\end{equation}
where $\bm f_{ba}$ is the force that magnetic dipole $\bmm_b$ exerts over dipole $\bmm_a$, $\bd$ is the vector between magnet centers ($d = |\bd|$) and $\mu_0$ is the value of the vacuum magnetic permeability. Here, the magnitude of the magnetic dipoles are considered equal,  i.e. $|\bmm_a|=|\bmm_b|=m$.

\begin{figure}[ht] 
	\centering \includegraphics[width=0.75\columnwidth]{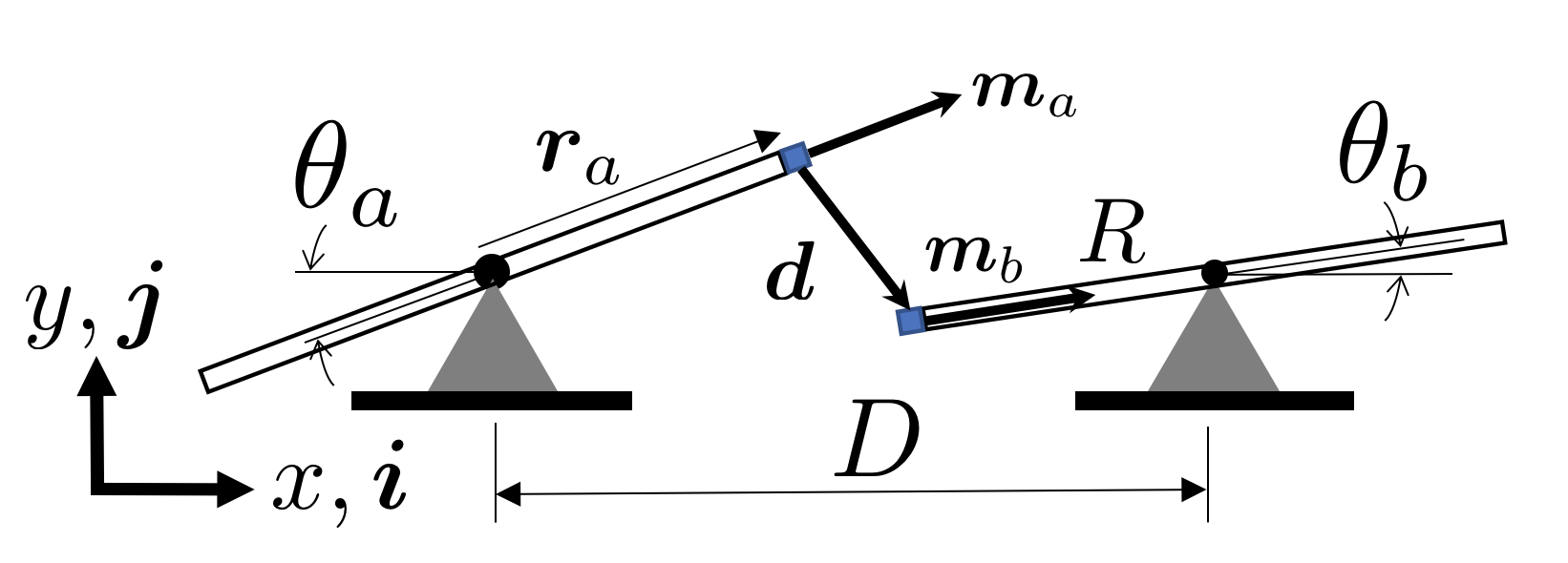}
	\caption{Sketch of two spinners interacting through permanent magnets: degrees of freedom and relevant parameters.}
	\label{Fig.Sketch.Magnets}
\end{figure}

According to the schematic of Fig.~\ref{Fig.Sketch.Magnets}, the dipole moments are expressed  as
\[\bmm_a=m  (\bm i \cos \theta_a + \bm j \sin\theta_a),\,\,\, \bmm_b=m  (\bm i \cos \theta_b + \bm j \sin\theta_b),\]
while the relative distance vector is 
\[\bd=\bm i \left[D - R \left( \cos\theta_a + \cos \theta_{b} \right)\right]-\bm j \left[R \left( \sin \theta_a + \sin \theta_b\right)\right]. \]

The interaction force can be conveniently resolved in terms of the unit vector pair $\bm i, \bm j$, i.e. $\bm f_{ab} =f_x \bm i+f_y \bm j$, where the two force components can be approximated through a Taylor series expansion about the equilibrium position $\theta_a,\theta_b\approx 0$. Truncation to the first order gives:
\begin{equation}
\begin{split}
&f_x= \frac{3m^2\mu_0}{2\pi\left( D-2R\right)^4 } + O\left( \theta_a^2,\theta_b^2,\theta_a\theta_b \right),\\
&f_y=-\frac{3m^2\mu_0(D+2R)}{4\pi\left( D-2R\right)^5 }\left( \theta_{a}+\theta_{b}\right)+ O\left( \theta_a^2,\theta_b^2,\theta_a\theta_b \right).
\end{split}
\end{equation}
The horizontal component is constant in linear regime, while the vertical one is proportional to the angle sum $(\theta_a+\theta_b)$, i.e. to the relative displacement between neighboring magnets in the vertical direction. The equation of motion for spinner $(a)$ is simply $I \ddot{\theta}_a-\mathcal{T}_{ba}(\theta_a,\theta_b) = 0$, and includes the moment corresponding to the interaction force, which is given by $\mathcal{T}_{ba}=|\br_a\times\bF_{ba}|$, where $\br_a=R  (\bm i \cos \theta_a + \bm j \sin\theta_a)$ is the vector that goes from center of spinner $a$ to the center of the magnet $\bm{m}_a$. This gives:
\begin{equation}
\begin{split}
\mathcal{T}_{ba}&=-\frac{3 m^2 \mu_0 R }{4 \pi (D-2 R)^5} (2(D-2 R)\theta_{a}\\
&+(D+2 R)( \theta_{b}+\theta_{a}))+O\left( {\theta_a}^3,{\theta_a}^2\theta_b,\theta_a{\theta_b}^2,{\theta_b}^3 \right)
\end{split}
\end{equation}

The expression above include one term depending solely on $\theta_a$ and another that is directly proportional to $(\theta_a+\theta_b)$.
The first term is analogous to the torque exerted by a spring connected to the ground, and is the result of the horizontal attractive force component between the magnets. 
The second term is proportional to the relative angular motion of neighboring spinners and is associated with the vertical component of the interaction force.

In order to account for nonlinearities in moderate rotation regimes, we extend the Taylor series expansion of the torque $\mathcal{T}_{ba}$ up to order 3, which gives:

\begin{widetext}
	\begin{equation}
	\begin{split}
	\mathcal{T}_{ba} = & -\frac{3 m^2 \mu_0 R }{4 \pi (D-2 R)^5}\Big(2(D-2 R)\theta_{a}+\left( D+2 R\right)\left( \theta_{b}+\theta_{a}\right)\Big)\\
	&+ \frac{m^2\mu_0R}{8 \pi (D-2 R)^7}\Big( (3D^3+12D^2R+3DR^2+16R^3)\left( \theta_{b}+\theta_{a}\right)^3\\
	& + (9D^3+4D^2R-46DR^2){\theta_{a}}^3
	+  (3D^3-6D^2R+42DR^2-96R^3) {\theta_{a}}^2\theta_{b}\;+ (6DR^2){\theta_{a}}{\theta_{b}}^2\\
	& + (-2D^3+10D^2R+2DR^2-32R^3){\theta_{b}}^3\Big) \; + O\left( {\theta_a}^4,...\right).
	\end{split} 
	\end{equation}
\end{widetext}

The nonlinear part of the torque includes five terms whose importance can be evaluated for the considered values of $D=70.9$ mm and $R=32.45$ mm, which gives $R/D\approx0.46$. Numerical estimation of the coefficients reveals that the term for $\left( \theta_{b}+\theta_{a}\right)^3$ is at least an order of magnitude larger than all other nonlinear coefficients. Therefore, the torque can be further approximated as follows:
\begin{equation}
\mathcal{T}_{ba}\approx - k_{\theta}  \theta_{a} - k_{t} \left( \theta_{a}+\theta_{b}\right) - \gamma \left( \theta_{a}+\theta_{b}\right)^3,
\end{equation}
where
\begin{equation}\label{Eq.coeffs}
\begin{split}
k_{\theta} &= \frac{6 m^2 \mu_0 R }{4 \pi (D-2 R)^5}(D-2 R)\\
k_{t} &= \frac{3 m^2 \mu_0 R }{4 \pi (D-2 R)^5}(D+2 R),\\
\gamma &= - \frac{m^2\mu_0R}{8 \pi (D-2 R)^7}(3D^3+12D^2R+3DR^2+16R^3).
\end{split}
\end{equation}
which leads to the following governing equation of motion for the spinner:
\begin{equation}
I \ddot{\theta}_a+k_{\theta,a}  \theta_{a} + k_{t,a} \left( \theta_{a}+\theta_{b}\right) + \gamma_a \left( \theta_{a}+\theta_{b}\right)^3= 0
\end{equation}

The negative sign in the nonlinear coefficient $\gamma$ in equation \eqref{Eq.coeffs} indicates that the cubic exponential term has a softening effect on the dynamic behavior of the spinner.

Please note that $k_\theta$ takes two different values in the chain $k_{\theta,a}$ and $k_{\theta,b}$ depending if the distance between spinners is $D_a$ or $D_b$ respectively. However, they add up in each spinner, since there is one spinner to the left and one to the right both contributing with a constant restoring longitudinal force $f_l$. As a result, all of them are the same $k_{\theta,a} + k_{\theta,b} = k_{\theta,b} + k_{\theta,a} = k_\theta$, except for three spinners: the left boundary $n=1$ is $k_{\theta,b}$, the right boundary $n=40$ is $k_{\theta,a}$, and the interface $n=21$ which is $2k_{\theta,a}$. This is taken into account in the analytic calculations.

Hence, the motion of regular $i$-th unit cell is expressed by Eq.~\eqref{Eq.Gov} and the motion of the inverted $i$-th unit cell is formulated as
\begin{widetext}
	\begin{equation}\label{Eq.Gov2}
	\begin{split}
	I \ddot{\theta}_{b,i}+k_{\theta}  \theta_{b,i} + k_{t,b}(\theta_{a,i}+\theta_{b,i})+k_{t,a} (\theta_{b,i}+\theta_{a,i-1})+\gamma_b(\theta_{a,i}+\theta_{b,i})^3+\gamma_a(\theta_{b,i}+\theta_{a,i-1})^3 & =  0\\
	I \ddot{\theta}_{a,i}+k_{\theta}  \theta_{a,i}+k_{t,a} (\theta_{b,i+1}+\theta_{a,i})+k_{t,b}(\theta_{a,i}+\theta_{b,i})+\gamma_a(\theta_{b,i+1}+\theta_{a,i})^3+\gamma_b(\theta_{a,i}+\theta_{b,i})^3 & = 0
	\end{split}
	\end{equation}
\end{widetext}

\section{Experimental evaluation of magnetic interaction coefficients}\label{Sec coefficients}
\subsection{Linear coefficients}\label{Sec linear coefficients}
\begin{figure}[ht] 
	\centering
	\begin{minipage}[]{1\linewidth}
		\subfigure[]{\includegraphics[width=0.8\textwidth]{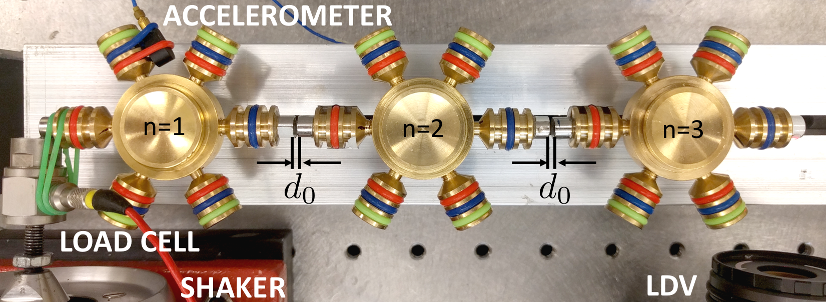}}
		\subfigure[]{\includegraphics[width=0.45\textwidth]{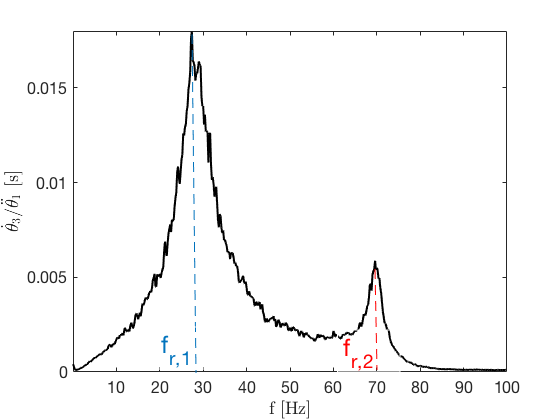}}
		\subfigure[]{\includegraphics[width=0.45\textwidth]{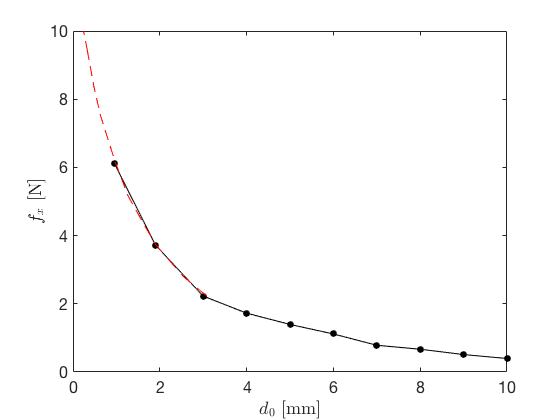}}
	\end{minipage}
	\caption{(a) Setup for the characterization of linear constants $k_\theta$ and $k_t$. The magnet distance is set at $d_0=1$ mm. (b) Frequency response function of the system showig the occurrence of two resonance frequencies that are related to the constants $k_\theta$ and $k_t$ and recorded for their estimation, which is based on repeating the meausrements for varying magnets distance $d_0$. (c) Comparison of the longitudinal attraction force $f_x$ evaluated on the basis of the estimated constants (black dots) and corresponding force provided in the technical specifications from the retailer (red dashed line).}
	\label{Fig.Characterization}
\end{figure}

The analytical model relies on the experimental estimation of linear and nonlinear coefficients $k_{\theta},\,k_t$ and $\gamma$ as a function of the distance between neighboring magnets faces $d_0=D-2R-h_m$, where $h_m=5$ mm is the height of the magnets.
To this end, we use a 3 spinner system which is tested dynamically. First, low-amplitude (linear) white noise excitation is applied to the left spinner $n=1$ in Fig. \ref{Fig.Characterization}a. The resonant frequencies of the resulting 2 degree of freedom system are recorded based on the evaluation of the response peaks. Estimation of the linear coefficients is based on the analytical expressions for these resonant frequencies, which are:
\[{f_{r_{1,2}}}^2=\frac{1}{2 \pi}\left( 3 k_{\theta} + 3 k_t \pm \sqrt{{k_{\theta}}^2 + 2 k_{\theta} k_t + 5 {k_t}^2}\right) /2 I\]
from which values of $k_{\theta}(d_0)$ and $k_t(d_0)$ are inferred. Examplary results are shown in Fig.~\ref{Fig.Characterization}b, while the full set of estimated coefficients are listed in Table \ref{Table.ExpValC1C2}. 

The estimated coefficients are subsequently used to evaluate the attractive horizontal component of the force $f_x(d_0)$, which is then compared with the data provided by the permanent magnets manufacturer (D4H2 nickel plated neodymium magnets by K\&J Magnetics, Inc.). The comparison in~Fig.~\ref{Fig.Characterization}c shows a very good agreeement and confirms the accuracy of the estimated coefficients, which are then used as inputs to the analytical model.

%\color{blue}
\def\arraystretch{1.2}%  1 is the default, change whatever you need
\begin{table}[hbtp]
	\begin{center}
		\caption{Experimental values of constants $k_{\theta}$ and $k_t$ as a function of distance between magnets $d_0=D-2R-h_m$.} % Note that the distance between magnet centers is $d_0+5\textrm{ mm}$.}
		\begin{tabular}{l  c c c c c c c}
			\hline
			$d_0$ (mm) & 1 & 2 & 3 & 4 & 5 & 6 & 7 \\
			\hline
			$k_{\theta}$ (Nm/rad) &  0.194 & 0.115 & 0.072 & 0.056 & 0.045 & 0.036 & 0.028 \\
			\hline
			$k_t$ (Nm/rad) & 2.385 & 1.224 & 0.720 & 0.406 & 0.282 & 0.178 & 0.127 \\
			\hline
		\end{tabular}
		\label{Table.ExpValC1C2}
	\end{center}
\end{table}
%\color{black}

\subsection{Nonlinear coefficients}\label{Sec nonlinear coefficients}
\begin{figure}[ht] 
	\centering
	\subfigure[]{\includegraphics[width=0.5\textwidth]{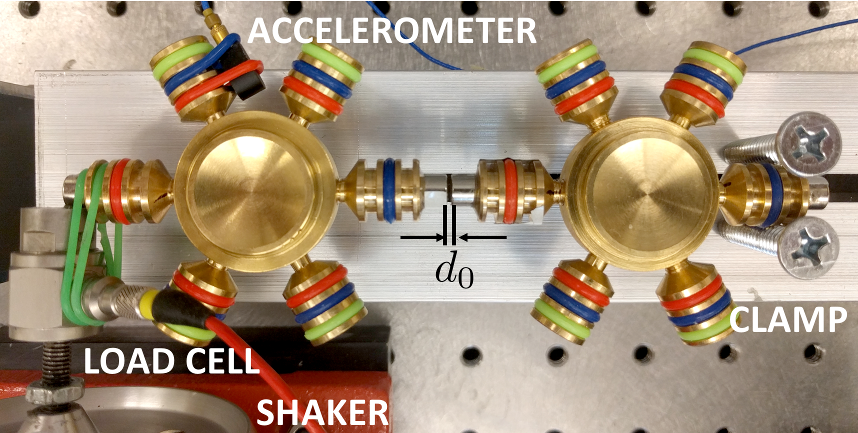}}\\
	\subfigure[]{\includegraphics[width=0.5\textwidth]{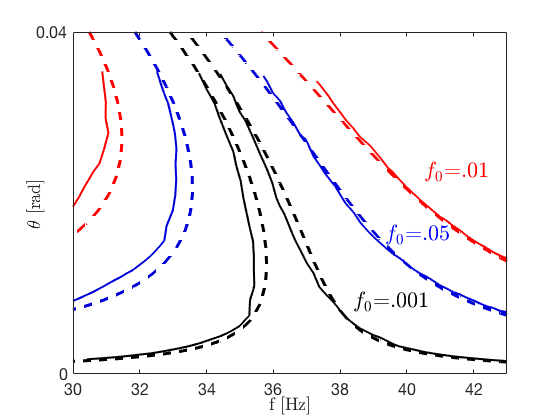}}
	\caption{(a) Setup for measuring the nonlinear forced response in a 1 dof system. Harmonic motion is imposed and the exerted force is measured for different amplitudes and frequencies. (b) Then the coefficient $\gamma$ is fine-tuned so that the analytical solution of the Duffing oscillator (dashed) matches the experimental results (solid) for different amplitudes of the force $f_0$.}
	\label{Fig.CharacterizationNonlinear}
\end{figure}

Subsequently, we estimate the nonlinear coefficient $\gamma$ using the 2-spinner system shown in Fig \ref{Fig.CharacterizationNonlinear}a. In this set-up, the left spinner $1$ is forced to oscillate harmonically at a particular amplitude and frequency, while spinner 2 is clamped in the $\theta_2=0$ position. We run a set of dynamic nonlinear steady-state experiments in which the exerted periodic force is recorded with a load cell (model 208C01 by PCB Piezotronics Inc.) from which the amplitude of its first harmonic $f_0$ is extracted.

Since the shaker is controlled in open-loop, we control the amplitude and frequency of the harmonic electronic signal sent to the shaker that imposes the motion $\theta_{1}=\Theta_1 e^{i\omega t}$, and its velocity is measured with the LDV, from which the amplitude of its first harmonic $\Theta_{1}$ is calculated. Then, for each experiment, we get a triplet of values: the amplitude of the response $\Theta_{1}$, its frequency, and the amplitude of the applied force $f_0$. The experiment is repeated over a range of imposed amplitudes from 0 to 0.04 rad and frequencies from 30 to 43 Hz. Mapping the results produces a surface that correlates frequency, amplitude of response and amplitude of applied force. The contours of this surface correlate frequency and amplitude of response for constant amplitude of excitation force $f_0$.

For this range of amplitudes and based on the assumptions described in Appendix \ref{Sec.AnalogMagnetic}, the governing equation of the forced response is equivalent to that of an undamped Duffing oscillator, 
\begin{equation}\label{Eq.DuffOscillator}
I \ddot{\theta_1} + \left( {k_{\theta}}+k_t\right) \theta_{1}+\gamma\theta_{1}^3=f(t)
\end{equation}
where $f(t)$ is the external force.

The response amplitude for harmonic excitation $f(t)=f_0 \cos(\omega t)$ can be obtained analytically from \cite{Nayfeh1993}
\[ \left(I\omega^2 -\left( k_\theta+k_t\right) -3/4\gamma A^2 \right)^2  A^2 = f_0^2\]
where $A$ is the amplitude of the response in $\theta_1=\Theta_1 e^{i\omega t}$, with $A=|\Theta_1|$. By comparing the measured response with the analytical predictions according to the expression above, we estimated a  value for the nonlinear coefficient equal to $\gamma=-320 \textrm{ Nm/rad}^3$ for a distance between neighboring magnets $d_0=1.2$ mm. The comparison is shown in Fig. \ref{Fig.CharacterizationNonlinear}d, which illustrates the excellent match between analytical predictions (dashed lines) and experimental results (solid lines) for the estimated value of $\gamma$. In the figure, each color relates amplitude and frequency for a different value of excitation force amplitude $f_0$. Through the same process, we estimate that $\gamma(d_0=1\textrm{ mm})=-366\textrm{ Nm/rad}^3$ and $\gamma(d_0=2\textrm{ mm})=-188\textrm{ Nm/rad}^3$.

%\onecolumngrid

\section{Experimental setup and methods}\label{Sec experimental set up}
\begin{figure}[ht] 
	\centering
	\includegraphics[width=0.99\textwidth]{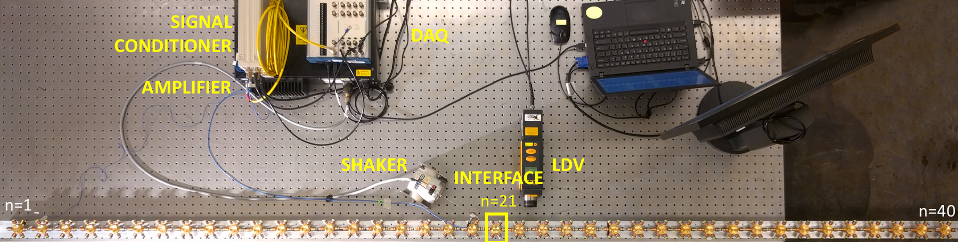}
	\caption{Physical 40 spinner system mounted on a beam. Distances between magnets are $d_{0,a}=1$ mm and $d_{0,2}=2$ mm. Transducers and data acquisition devices are also shown.}
	\label{Fig.Setup2}
\end{figure}

The complete spinner chain is bolted to a straight slotted beam, which allows adjusting the inter-magnetic distances as needed by the experiments. The spinner radius is $R=32.45$ mm, which leads to a rotary inertia value of $I=37.2 \textrm{ Kg mm}^2 $ including the magnets. We calculate the inertia using meticulous measurements of the volumes and masses of all the parts conforming each spinner. We measured all the geometries in spinner: the main body, the pegs, the bearings and the bearing balls. Then we created a detailed CAD model and calculated the volumetric inertias ($m^5$) of three different parts: the spinner body and pegs, the bearing outer cylinder, and the bearing balls. Those were obtained by numerical integration about the axis of gyration $I_V=\sum\left( r_i^2\delta V_i\right) $, where $r_i$ is the distance between the center of the $i$-th differential volume $\delta V_i$ and the axis of gyration. We also calculated the volumes and weighted the parts separately. Assuming that the materials are homogeneous, we estimate the densities $rho_m$ of each part $m$. We calculate the mass inertia by multiplying the volumetric inertia by the density of each part $I=\sum I_{V,m} \rho_m$. The bearing balls contribute half because its motion is half of the rest of the spinner. We neglected the spinning of the bearing balls in the motion. The magnets, which are $5$ mm tall and $6.35$ mm diameter, are placed at distances $d_{0,a}=1$ mm and $d_{0,b}=2$ mm apart. The corresponding distances between the centers of the spinners are respectively $D_a=70.9$ mm and $D_b=71.9$ mm. Figure~\ref{Fig.Setup2} shows a top view of the experimentally tested 40 spinner chain.

In the experiments we impose harmonic motion to the spinner at the left boundary $\theta_1$ or the spinner next to the interface $\theta_{20}$ depending on each experiment goal, with a shaker controlled in open loop. The shaker, a model V201 by LDS LTD., is excited with an electronic signal programmed in the PC and sent through the data acquisition system (DAQ), (USB-6366 782263-01 by National Instruments TM). We measure the acceleration of the excited spinner using the accelerometer (model 352A24 by PCB Piezotronics Inc.) and calculate its motion by integration. The motion of the other spinners is calculated from integration of the velocities, which in turn are measured by LDV using a PDV-100 scanning head by Polytech GmbH. This is a single point LDV, so we repeat the experiments 40 times and move the LDV device manually between locations to measure the motion of all the spinners. The DAQ is used to trigger the excitations and measurements always with the same time interval between them, which ensures that the steady-state is reached and that phase is synchronized between experiments. 

The signal imposed to the shaker is either white noise over the frequency range of interest ($0-80$ Hz) to provide the response of the system in the frequency domain, or harmonic for steady-state measurements. The signal is properly amplified to obtain the targeted amplitudes of displacement in the shaker. These amplitudes are monotonically but not proportionally related to the amplitude of the electronic signal that excites the shaker. Therefore, we can increase and decrease the amplitude of motion $\Theta_{20}$ imposed to spinner $n=20$ without knowing its exact value \textit{a priori}. The exact value of the motion is calculated \textit{a posteriori} from the accelerometer measurements. At the same time, the force at the shaker tip is measured using a force transducer model 208C01 by PCB Piezotronics Inc. These signals are amplified for acquisition using a signal conditioner model 482A21 by PCB Piezotronics Inc.

Finally, videos of the motion in the steady-state nonlinear experiments are recorded using a high speed camera model 675K-M1 by Photron USA, Inc. placed right above the spinners system (not shown in the figure). Due to the length of the chain, all the 40 spinners do not fit in the camera frame if we want to maintain a good level of resolution. Therefore, we use 15 different camera positions, recording 2 or 3 spinners at a time. We use the DAQ to control and coordinate the excitation, the measurements and the camera trigger, so that we ensure phase synchronization between the videos. These were later postprocessed and stitched together using Matlab software.

The snapshots of the deformed configurations of the chain shown in Fig.~\ref{Fig snapshots} are extracted from the movies provided as supplementary material. In the snapshots and in the movies, visualization of the angular rotation of the spinners is aided by superimposing to each spinner a colored circle of radius proportional to the amplitude of motion. Also, the rotation angle is extracted from the video by employing in-house Digitial Image Correlation software. The lengthwise variation of the rotation angle of the spinners is shown in the graphs accompanying each of the response movie, which helps visualizing the spatial extent of motion and differentiating localized modes versus bulk-propagating modes. 

In detail, we provide the following movies as supplementary material:

\begin{itemize}
	
	\item[SM1] Description of the experimental set-up and animation explaining the spinner lattice visualization in Figure\ref{Fig snapshots}. The experiments are conducted by repeating the measurements over 15 separate portions of the lattice, as the entire length exceeds the aperture of the camera. Upon recording, the measurements are phase-matched and stitched to obtain a single recording for an assigned amplitude of motion.
	
	\item[SM2] Experimental results recorded for low amplitude excitation, and corresponding to the still picture of Fig. ~\ref{Fig snapshots}.a. The recorded video data are used to extract angular information about the rotation of the spinners, which is plotted as a function of the spinner number in the bottom graph, which helps visualising the localized nature of the dynamic deformed lattice response at low amplitude excitation, which corresponds to the induced edge mode.
	
	\item[SM3] Experimental results recorded for medium amplitude excitation, and corresponding to the still picture in Fig. \ref{Fig snapshots}.b. The plot of the angular motion of the spinners shows the increase in penetration of the dynamic response which extends away from the interface as amplitude increases. 
	
	\item[SM4] Experimental results recorded for high amplitude excitation, and corresponding to the still picture of Fig.~\ref{Fig snapshots}.c. The plot of the spinners' rotation clearly shows that the mode now extends to the entire length of the chain.
\end{itemize}

%%%%%%%%%%%%%%%%%%%%%%%%%%%%%%%%%%%%%%%

\end{document}